\documentclass[aps,pra,twocolumn,groupedaddress,amsmath,amssymb,10pt, showpacs]{revtex4-1}
\usepackage{graphicx}% Include figure files
\usepackage{bm}% bold math
\usepackage{hyperref}
\usepackage{graphicx,dsfont,amsthm}
\newcommand{\qudit}[1]{\left\vert #1 \right\rangle}
\newcommand{\rqudit}[1]{\left\langle #1 \right\vert}

\newcommand{\Z}{\mathbb{Z}}

\newcommand{\I}{\mathbb{I}}
\newcommand{\C}{\mathbb{C}}
\newtheorem{theorem}{Theorem}

\newtheorem{remark}{Remark}

\begin{document}
\title{Deriving a basis for a set of bounded operators on the Hilbert space ${\C}^d$}

\author{Colin M.~Wilmott}
\email{wilmott@thphy.uni-duesseldorf.de}

\affiliation{Institute for Theoretical Physics III,
Heinrich-Heine-Universit\"at D\"usseldorf, 40225 D\"usseldorf,
Germany. }
\begin{abstract}
Of crucial importance to the development of quantum computing and
information has been the construction of a quantum operations
formalism that admits a description of quantum noise while
simultaneously revealing the behavior of an open quantum system.
The operator-sum representation is such a formalism and has
provided a succinct description for set of bounded operators that
act on a finite dimensional quantum system. In this paper we
derive a basis for the set of bounded operators that act on a
$d$-dimensional Hilbert space and  we illustrate how this basis
set may be extended and identified with a set of elements upon
which the operator-sum representation rests.
\end{abstract}

\maketitle
\section{Introduction}

The desire to comprehend philosophies at the edge of possibility
continues to be a source of intellectual advancement today as it
has been at any other time in history. If such a premise is taken
to be a departure point in the challenge to extend the boundaries
of knowledge then theoretical and technical innovation will abide.
Indeed, the theory of quantum information has advanced the concept
of information established in the seminal work of Claude Shannon
\cite{shannon} and has inspired  discoveries whose very nature lie at
the frontier of reality.

Modern computing rests with the pioneering work of Charles Babbage
\cite{babbage} and Alan Turing \cite{turing}. An analytical
machine put forward by Babbage conceived the principle on which
modern computing rests. Over a century later, Turing improved the
ideas of Babbage by devising a programmable means that would
become the basis for computing logic. However, challenges from a
new information paradigm emerged with Richard Feynman's promise of
quantum computing \cite{feynman}. In 1985 David Deutsch
\cite{deutsch} gave credence to Feynman's proposition by
demonstrating the principle on which a quantum computer may be
constructed. Quantum mechanics now offered the theory of
computation a new direction with profound consequences.

The construction of a quantum computer is predicated on realizing
the inherent computational processing advantage of quantum
computation over its classical analogue. Fundamentally, the real
power of a quantum computer rests with an ability to control
quantum interference effects while utilizing the inherent
principle of parallelism \cite{deutsch}. Nevertheless, quantum
computing comes with a cost and this is best explained by a set of
interactions with an open quantum system called the
\emph{environment}. It is through these interactions that quantum
noise is introduced into the quantum system  leading to the
process of \emph{decoherence} and the complete corruption of
quantum information. While explaining an exact model of quantum
system-environment interaction is difficult, considerable
theoretic insights into quantum noise and its effects on quantum
systems have been made by studying properties of this interaction
\cite{kraus}. In particular, the \emph{operator-sum
representation} \cite{kraus, nielsenchuang00} has provided a
descriptive process of quantum system-environment interactions
while capturing the dynamic change to a quantum state as a result
of a quantum process \cite{nielsenchuang00}.

In this paper we propose to better  understand the fundamentals of noise in a quantum system.
We will derive a basis for the set of bounded
operators that act on a $d$-dimensional Hilbert space, and given
basic initial assumptions, we will illustrate how this basis
set may be extended to identify a set of operational elements that
coincide with the operator-sum representation for a quantum operation.

\section{Deriving a basis for a set of bounded operator on $\C^d$}\label{sec2}
The challenge of quantum computing and information is to elicit a
reliable form of communication and to maintain such a form in the
presence of quantum noise. Noise is an inevitable characteristic
that subjects a quantum state to unwanted interactions with the
environment. Any strategy to mitigate against quantum noise
ultimately requires an understanding of the quantum operation
process. To this end, the operator-sum representation \cite{kraus,
nielsenchuang00} has provided a succinct description for the set
of quantum operations that act on a quantum state thereby allowing
us to understand quantum noise and its effects. We shall now
characterize the dynamics of a quantum operation before deriving a
set of elements that constitute a basis set for an operator-sum
representation of a quantum operation.
\subsection{Preliminaries}\label{pre}
Consider a $d$-dimensional Hilbert space $\C^d$ and fix each basis
state  to correspond to an element of ring $\Z_d$ of integers
modulo $d$. We shall regard the Hilbert space $\C^d$ as the
\emph{principal quantum system}. The basis $\{\qudit{i}, \ i \in
\Z_d\} \subset \C^d$ whose elements correspond to the column
vectors of the identity matrix ${\I}$ is called the computational
basis. A \emph{qudit} is then a $d$-dimensional quantum state
$\qudit{\psi} \in {\C^d}$ written as $\qudit{\psi} =
\sum^{d-1}_{i=0}{\alpha_i\qudit{i}}$ where $\alpha_i \in \C$ and
$\sum^{d-1}_{i=0}{\vert\alpha_i\vert^2} = 1$.

To every Hilbert space $\C^d$ there corresponds a \emph{dual
space} $({\C^d})^{\perp}$ consisting of the set of all linear
functionals  from $\C^{d}$ to the complex numbers. In particular,
if $\{\qudit{i},\ i \in \Z_d\}$ is a basis for $\C^d$, then there
is a uniquely determined basis $\{\qudit{j}^{\dagger},\ j \in
\Z_d\}$ in $({\C^d})^{\perp}$ such that the linear functional
$j(\qudit{i})$ is identically ${\qudit{j}}^{\dagger}(\qudit{i}) =
\delta_{i,j}$. In the language of Dirac, the action of the
conjugate linear map $j(\qudit{i})$ on $\C^{d}$ is written and
defined as $\langle j\vert i\rangle = \delta_{i,j}$ with
$\delta_{i,j}$ denoting the Kronecker delta.

Finally, given $\C^d$ we have with us the space of linear
bounded operators on $\C^d$. The study of such operators is
known to elicit a matrix representation. Therefore, supposing that $\qudit{\psi}$ and
$\qudit{\phi}$ are states in the Hilbert space $\C^{d}$, we define
$\qudit{\phi}\rqudit{\psi}$ to be the linear operator on $\C^{d}$
that maps $\qudit{\psi}$ to $\qudit{\phi}$.

\subsection{Deriving a space of bounded operators on $\C^d$}\label{pre}

Consider the quantum state in a principal system $\qudit{\psi} \in
\mathcal \C^d$, and further consider an adjoined environment state
$\qudit{E}$ endowed with an orthonormal basis of dimension $d^2$.
Since $\C^d$ is a $d$-dimensional Hilbert space, it suffices that
the environment be a $d^2$-dimensional Hilbert space
\cite{nielsenchuang00}. Without loss of generality, we suppose the
environment state to be initially a pure state. We assume that the
principal system and the environment are initially uncorrelated
with the joint state of the system-environment then given as the
product state $\qudit{\psi} \otimes \qudit{E}$.

Our goal will be to chart the dynamics of the principal system and
environment by allowing the joint state $\qudit{\psi} \otimes
\qudit{E}$ to evolve according to some unitary operation. As a
consequence, we will explicitly derive a basis for the set of
operators that act on a single qudit state.

Let $U$ be  a unitary operation acting on the product state
$\qudit{\psi} \otimes \qudit{E}$, and let us write the interaction
of each basis qudit with the environment under $U$ as
\begin{eqnarray}
U(\qudit{i} \otimes \qudit{E}) &=&
\sum_{l=0}^{d-1}{}\gamma_{-i+l,-i}(\qudit{i+l}\otimes
\qudit{e_{-i+l,-i}})\nonumber\\&=&
\sum_{l=0}^{d-1}{}\qudit{i+l}\otimes
\gamma_{-i+l,-i}\qudit{e_{-i+l,-i}},\end{eqnarray} for $i \in
\{0,\dots,d-1\}$. By linearity of quantum mechanics, we have it
that the action of the unitary operation  $U$ on the product state
$\qudit{\psi} \otimes \qudit{E}$ may then be written as
\begin{eqnarray}
U(\qudit{\psi} \otimes
\qudit{E})&=&U\left(\left(\sum_{i=0}^{d-1}{}\alpha_i\qudit{i}\right)\otimes
\qudit{E}\right)\nonumber\\
&=& U\left(\sum_{i=0}^{d-1}{}\alpha_i(\qudit{i}\otimes
\qudit{E})\right)\nonumber\\
&=& \sum_{i=0}^{d-1}{}\alpha_iU(\qudit{i}\otimes
\qudit{E})\nonumber\\
&=& \sum_{i=0}^{d-1}{}\sum_{l=0}^{d-1}{}\alpha_i\qudit{i+l}
\otimes \gamma_{-i+l,-i}\qudit{e_{-i+l,-i}}.\ \ \ \ \ \
\label{env}\end{eqnarray}
\\
Let $\omega$ be a primitive $d^{\textrm{th}}$ root of unity such
that $\omega^d =1$ and $\omega^d \ne 1$ for all $0 < t < d$. Using
the fact that $\frac{1}{d}\sum_{z=0}^{d-1}{}\omega^{zk} = 1$ if
$k=0$ and vanishes otherwise, we have it that Eq.~(\ref{env}) may
be written as
\begin{widetext}
\begin{eqnarray}\label{three}
&&\sum_{i=0}^{d-1}{}\sum_{l=0}^{d-1}{}\alpha_i\qudit{i+l}\otimes
\gamma_{-i+l,-i}\qudit{e_{-i+l,-i}}\nonumber\\
&& =  \frac{1}{d}\sum_{i=0}^{d-1}{}\sum_{l=0}^{d-1}{}\left(\alpha_i\qudit{i+l} \otimes\left( \sum_{z=0}^{d-1}{}\sum_{k=0}^{d-1}{}\omega^{zk}\gamma_{-i+l+z,-i+z}\qudit{e_{-i+l+z,-i+z}}\right)\right)\nonumber\\
&&  = \frac{1}{d}\sum_{i=0}^{d-1}{}\sum_{l=0}^{d-1}{}\sum_{k=0}^{d-1}{}\left(\alpha_i\qudit{i+l}\otimes \left(\sum_{z=0}^{d-1}{}\omega^{zk}\gamma_{-i+l+z,-i+z}\qudit{e_{-i+l+z,-i+z}}\right)\right)\nonumber\\
&&  = \frac{1}{d}\sum_{l=0}^{d-1}{}\sum_{k=0}^{d-1}{}\left(\sum_{i=0}^{d-1}{}\left(\alpha_i\qudit{i+l}\otimes \left( \sum_{z=0}^{d-1}{}\omega^{zk}\gamma_{-i+l+z,-i+z}\qudit{e_{-i+l+z,-i+z}}\right)\right)\right)\nonumber\\
&& =
\frac{1}{d}\sum_{l=0}^{d-1}{}\sum_{k=0}^{d-1}{}\left(\sum_{i=0}^{d-1}{}\left(\omega^{ik}\alpha_i\qudit{i+l}\otimes
\left(\sum_{z=0}^{d-1}{}\omega^{-ik}\omega^{zk}\gamma_{-i+l+z,-i+z}\qudit{e_{-i+l+z,-i+z}}\right)\right)\right)\nonumber\\
&& = \frac{1}{d}\sum_{l=0}^{d-1}{}\sum_{k=0}^{d-1}{}\left(\sum_{i=0}^{d-1}{}\left(\omega^{ik}\alpha_i\qudit{i+l}\otimes
\left(\sum_{z'=0}^{d-1}{}\omega
^{z'k}\gamma_{z'+l,z'}\qudit{e_{z'+l,z'}}\right)\right)\right)\nonumber
\end{eqnarray} \begin{eqnarray}
&& =
\frac{1}{d}\sum_{l=0}^{d-1}{}\sum_{k=0}^{d-1}{}\left(\left(\sum_{i=0}^{d-1}{}
\omega^{ik}\alpha_i\qudit{i+l}\right)\otimes
\left(\sum_{z'=0}^{d-1}{}\omega^{z'k}\gamma_{z'+l,z'}\qudit{e_{z'+l,z'}}\right)\right).
\end{eqnarray}
\end{widetext}

The linear product representation now describes the set of
operators that act on the product state $\qudit{\psi} \otimes
\qudit{E}$ with respect to the unitary operation $U$. In
particular, the operator $X_1 =
\sum_{i=0}^{d-1}{}\qudit{i+1}\rqudit{i}$ maps $\alpha_i\qudit{i}$
to $\alpha_i\qudit{i+1}$ for $i \in \{\qudit{0},
\dots,\qudit{d-1}\}$, and thus maps
$\sum_{i=0}^{d-1}{}\alpha_i\qudit{i}$ to
$\sum_{i=0}^{d-1}{}\alpha_i\qudit{i+1}$. Similarly, the operator
$Z_1 = \sum_{i=0}^{d-1}{}\omega^i\qudit{i}\rqudit{i}$ maps
$\sum_{i=0}^{d-1}{}\alpha_i\qudit{i}$ to
$\sum_{i=0}^{d-1}{}\omega^i\alpha_i\qudit{i}$. When taken together
 both $X_1$ and $Z_1$ are called the \emph{Weyl Pair} \cite{Weyl31}.
In rewriting Eq.~(\ref{three}), we have, up to a scalar,
\begin{eqnarray}
&&\sum_{l=0}^{d-1}{}\sum_{k=0}^{d-1}{}\left(\sum_{i=0}^{d-1}{}\omega^{ik}\alpha_i\qudit{i+l}\right)\otimes
\left(\sum_{z'=0}^{d-1}{}\omega^{z'k}\gamma_{z'+l,z'}\qudit{e_{z'+l,z'}}\right)\nonumber\\
%\label{construction}
&& = \sum_{l=0}^{d-1}{}\sum_{k=0}^{d-1}{}X_lZ_k\qudit{\psi}
\otimes \gamma_{lk}\qudit{e_{lk}}\label{XZ}.
\end{eqnarray}
Thus, the action of the unitary operation $U$ on the joint state
$\qudit{\psi}\otimes\qudit{E}$ yields the set of operators
$\{X_lZ_k = \sum_{i=0}^{d-1}{}\omega^{ik}\qudit{i+l}\rqudit{i}, \
(l,k) \in \Z_{d}\times\Z_{d}\}$ that act on the principal system.

We now  show that the derived set  $\{X_lZ_k, \ (l,k) \in \Z_d
\times \Z_d\}$ indeed forms a basis for a space of bounded
operators on $\C^d$.
\begin{theorem}
\label{niceproof} Let $\omega$ be the primitive $d^{\textrm{th}}$
root of unity and let $X_{l}\vert i \rangle = \qudit{i + l\
(\textrm{mod}\ {d})}$ and $Z_{k}\qudit{i} = \omega^{ik}\qudit{i}$.
Then, the set  $\left\{ X_{l}Z_{k}, \ (l,k)\in \Z_{d} \times
\Z_{d}\right\}$ forms a basis for the $d$-dimensional Hilbert
space $\C^d$.
\end{theorem}
\begin{proof} To show that the set $\{X_lZ_k, \ (l,k) \in \Z_d
\times \Z_d\}$ is linearly independent and spans $\C^d$, it
suffices to show that the basis $\{\qudit{a}\rqudit{b}, \ a,b \in
\Z_{d}\}$ for $\C^d$ may be written as a linear combination of
elements from $\{X_lZ_k, \ (l,k) \in \Z_d \times \Z_d\}$. This
follows since both of these sets have cardinality $d^2$. Let us
consider the set $\{X_lZ_k, \ (l,k) \in \Z_d \times \Z_d\}$ in the
$\{\qudit{a}\rqudit{b}, \ a,b \in \Z_{d}\}$ basis as
\begin{eqnarray}
X_lZ_k = \sum_{i=0}^{d-1}{}\omega^{ik}\qudit{i+l}\rqudit{i},
\end{eqnarray}
for $k,l \in \Z_d$. Then $X_lZ_k\qudit{i} =
X_l\omega^{ik}\qudit{i} = \omega^{ik}\qudit{i+l}$. Now, let us
suppose that $\qudit{a}\rqudit{b}$ is expressed as the linear
combination $\qudit{a}\rqudit{b} = \sum_{(l,k) \in
\Z_{d}\times\Z_{d}}{\mathcal{\xi}}_{l,k}X_lZ_k$. The coefficients
${\xi}_{l,k}$ are given by
\begin{eqnarray}
{\mathcal{\xi}}_{l,k} &=&
\frac{1}{d}\textrm{tr}\left((X_lZ_k)^{\dagger}\qudit{a}\rqudit{b}\right)\nonumber\\
&=&
\frac{1}{d}\textrm{tr}\left(\sum_{i=0}^{d-1}\omega^{-ik}\qudit{i}\rqudit{i+l}{a}\rangle\rqudit{b}\right)\nonumber\\
&=&  \frac{1}{d}\ \omega^{-bk}\  \langle b+l \vert a
\rangle,\nonumber\end{eqnarray}\begin{eqnarray}
 &=&  \frac{1}{d}\ \omega^{-bk}\ \delta_{b+l,a},
\end{eqnarray}
where
$\delta_{l,k}$ is the Kronecker delta; \begin{eqnarray}\delta_{l,k} = \left\{%
\begin{array}{ll}
    1, & \hbox{\textrm{if} \ $l = k$} \\
    0, & \hbox{\textrm{if} \ $l \ne k$.}\\
\end{array}%
\right. \end{eqnarray} We show that with $\xi_{lk}$ defined as
these values then $\qudit{a}\rqudit{b}$ is in the span of
$\{X_lZ_k, \ (l,k) \in \Z_d \times \Z_d\}$. Now,
\begin{eqnarray}
&&(X_{l}Z_{k})^{\dagger}\qudit{a}\rqudit{b}\nonumber\\
&& = (X_{l}Z_{k})^{\dagger}\sum_{(m,n) \in
\Z_{d}\times\Z_{d}}{}\xi_{m,n}X_{m}Z_{n}\nonumber\\
&& = \xi_{l,k}I + \sum_{(m,n) \ne
(l,k)}{}\xi_{m,n}(X_{l}Z_{k})^{\dagger}X_{m}Z_{n}
\end{eqnarray}
with $(X_{l}Z_{k})^{\dagger}X_{m}Z_{n}$ possessing a vanishing
trace. Since
\begin{eqnarray}
&&\sum_{(l,k) \in
\Z_{d}\times\Z_{d}}{}\frac{1}{d}\omega^{-bk}\delta_{b+l,a}X_{l}Z_{k}\nonumber\\
&&  \ =  \sum_{(l,k) \in
\Z_{d}\times\Z_{d}}{}\frac{1}{d}\omega^{-bk}\delta_{b+l,a}\left(\sum_{i=0}^{d-1}{}\omega^{ik}\qudit{l+l}\rqudit{i}\right)\nonumber\\
&&  \ = \sum_{i=0}^{d-1}{}\sum_{(l,k) \in \Z_{d}\times\Z_{d}}{}\frac{1}{d}\omega^{(i-b)k}\delta_{b+l,a}\qudit{i+l}\rqudit{i}\nonumber\\
&& \ =  \sum_{i=0}^{d-1}{}\sum_{k=0}^{d-1}{}\frac{1}{d}\omega^{(i-b)k}\qudit{i+a-b}\rqudit{i}\nonumber\\
&& \ = \sum_{i=0}^{d-1}{}\delta_{i,b}\qudit{i+a-b}\rqudit{i}\nonumber\\
 &&  \ = \qudit{a}\rqudit{b}
\end{eqnarray}
as   $\sum_{k=0}^{d-1}{}\frac{1}{d}\omega^{(i-b)k} =
\delta_{i,b}$, we then have it that  $\langle b \vert \sum_{(l,k)
\in \Z_{d}\times\Z_{d}} {} {\xi}_{l,k}X_{l}Z_{k}\vert a \rangle$ $
= \delta_{a,b}$. Finally, $\qudit{a}\rqudit{b} = \sum_{(l,k) \in
\Z_{d}\times\Z_{d}}{} {\xi}_{l,k}X_{l}Z_{k}$, and the result
follows. \end{proof}

\begin{remark} The set $\{X_lZ_k, \ (l,k) \in \Z_d \times \Z_d\}$ forms
a $d^2$-dimensional Lie algebra with a $d \times d$ matrix
representation defined by the matrices with entries {{
\textsf{X}}}$_{m,n} = \delta_{m,n-l}$, {{\textsf{Z}}}$_{m,n} =
\omega^{mk}\delta_{m,n}$.
\end{remark}

We have derived a set of linear operators that describe the action
of a quantum operator on a principal qudit system and we have also
shown that this derived set forms a basis for the space of
operators on $\C^d$. Finally, we note that Eq.~\ref{XZ} provides a
conceptual foundation for many aspects of quantum information.

\section{On the operator-sum representation}

The quantum operations formalism is a general description for the
evolution of a quantum system. A more succinct description for
this formalism is given by the operator-sum representation,
see \cite{nielsenchuang00} and references therein. The
operator-sum representation is a significant quantum theoretical
model in that it  concisely characterizes the set of changes that arise when a
quantum system evolves in time. We shall now outline  the operator-sum representation for a quantum operation $\cal E$.

We begin with a principal system to which we adjoin an
environment system which we  assume is spanned by  an orthonormal set of
basis states $\qudit{e_m}$,  $1\leq m \leq  d^2$. We suppose that
environment is prepared in the pure state
$\qudit{e_0}\rqudit{e_0}$ and  that the principal system and
environment are initially uncorrelated. We  further suppose  the state
of the joint system may be written  as the product state
$\rho\ \otimes\qudit{e_0}\rqudit{e_0}$. Next, we apply a unitary
operation $U$ to the joint state before implementing a partial trace
over the environment. We summarize the operator-sum representation
of a quantum operation $\cal E$ on a state $\rho$ as
\begin{eqnarray}
{\cal E}(\rho) &=& {\textrm{tr}_{\textrm{env}}}
\left[U(\rho\otimes\qudit{e_0}\rqudit{e_0})U^\dagger\right] \nonumber\\ &=&
\sum_{m}{}\rqudit{e_m}U(\rho\otimes\qudit{e_0}\rqudit{e_0})U^\dagger\qudit{e_m}\nonumber\\
&=& \sum_m{}E_m\rho E_m^\dagger\label{operator-sum}.
\end{eqnarray}
The set $\{E_m, \  1\leq m \leq  d^2\}$ represents the set of
operation elements for the quantum operation $\cal E$ and
satisfies the completeness relation $\sum_k{}E_mE_m^\dagger = I$.

Now, let us recall Sec.~\ref{pre} wherein we considered a pure
state of a principal system adjoined an
environment system spanned by an orthonormal basis set of
dimension $d^2$. We defined a unitary operator $U$ to act on the
state $\qudit{\psi} \otimes{\qudit{E}}$ of the joint system.  The
result of the unitary operator $U$ on
$\qudit{\psi}\otimes{\qudit{E}}$ yielded a set of operators acting
on the principal system;
\begin{eqnarray}
U(\qudit{\psi}\otimes{\qudit{E}}) &=&
\frac{1}{d}\sum_{l=0}^{d-1}{}\sum_{k=0}^{d-1}{}X_lZ_k\qudit{\psi}
\otimes \gamma_{lk}\qudit{e_{lk}}.
\end{eqnarray}

Motivated by a question raised by Nielsen and Chuang
\cite{nielsenchuang00} and by the material of Sec.~\ref{pre}, we
now consider the question of how an operator-sum representation
can be determined for an open quantum system. To this end, we will
extend the unitary operator $U$ defined in Sec.~\ref{pre} to act on the
entire state space of the joint system. The representative input state is $\rho \otimes \qudit{e_0}\rqudit{e_0}$ and is written
\begin{eqnarray}
&&\rho \otimes \qudit{e_0}\rqudit{e_0}\nonumber\\
&& \quad = \sum_{s\in\vert
S\vert}^{}{}\sum_{i=0}^{d-1}{}\sum_{j=0}^{d-1}{}p_s\alpha^s_{ij}\qudit{i}\rqudit{j}\otimes\qudit{e_o}\rqudit{e_0}
\end{eqnarray}
with $p_s\geq0$ and $\sum_{s\in \vert S\vert}{p_s} = 1.$ Next, we allow the unitary operator
$U$ to act on the initial state $\rho \otimes \qudit{e_0}\rqudit{e_0}$ according to $U(\rho
\otimes \qudit{e_0}\rqudit{e_0})U^\dagger$. By linearity, we have  it  that
\begin{widetext}
\begin{eqnarray}
&& U\left(\rho \otimes \rho_{env}\right)U^{\dagger} \nonumber\\
&&  = U\left(\left(\sum_{s\in\vert S\vert}^{}{}\sum_{i=0}^{d-1}{}\sum_{j=0}^{d-1}{}p_s\alpha^s_{ij}\qudit{i}\rqudit{j}\right)\otimes\qudit{e_o}\rqudit{e_0}\right)U^{\dagger}\nonumber\\
&&  = U\left(\sum_{s\in\vert S\vert}^{}{}\sum_{i=0}^{d-1}{}\sum_{j=0}^{d-1}{}p_s\alpha^s_{ij}\left(\qudit{i}\rqudit{j}\otimes\qudit{e_o}\rqudit{e_0}\right)\right)U^{\dagger}\nonumber\\
&&  = \sum_{s\in\vert S\vert}^{}{}\sum_{i=0}^{d-1}{}\sum_{j=0}^{d-1}{}p_s\alpha^s_{ij}U\left(\qudit{i}\rqudit{j}\otimes\qudit{e_o}\rqudit{e_0}\right)U^{\dagger}\nonumber\\
&&  = \sum_{s\in\vert S\vert}^{}{}\sum_{i=0}^{d-1}{}\sum_{j=0}^{d-1}{}\sum_{l=0}^{d-1}{}p_s\alpha^s_{ij}
\qudit{i+l}\rqudit{j+l}\otimes\gamma_{-i+l,-i}\gamma^*_{j,j+l}\qudit{e_{-i+l,-i}}\rqudit{e_{j,j+l}}.\label{operator}
\end{eqnarray}\end{widetext}
We shall now make repeated use of an earlier stated fact. Let $\omega$ be a
primitive $d^{\textrm{th}}$ root of unity. It then follows that
$\frac{1}{d}\sum_{z=0}^{d-1}{}\omega^{zk} = 1$ when $k=0$ while
$\frac{1}{d}\sum_{z=0}^{d-1}{}\omega^{zk}$ vanishes in all other
instances. In rewriting Eq.~\ref{operator}, we have
\begin{widetext}
\begin{eqnarray}
&&\sum_{s\in\vert S\vert}^{}{}\sum_{i=0}^{d-1}{}\sum_{j=0}^{d-1}{}\sum_{l=0}^{d-1}{}p_s\alpha^s_{ij}
\qudit{i+l}\rqudit{j+l}\otimes\gamma_{-i+l,-i}\gamma^*_{j,j+l}\qudit{e_{-i+l,-i}}\rqudit{e_{j,j+l}}\nonumber\\
&&\begin{split}
&  = \frac{1}{d^2}\sum_{s\in\vert S\vert}^{}{}\sum_{i=0}^{d-1}{}\sum_{j=0}^{d-1}{}\sum_{l=0}^{d-1}{}\Biggl(p_s\alpha^s_{ij}
\qudit{i+l}\rqudit{j+l}\nonumber\\
 &\quad\otimes\Biggl(\sum_{k=0}^{d-1}{}\sum_{z_1=0}^{d-1}{}
\sum_{z_2=0}^{d-1}{}\omega^{z_1k}\omega^{z_2k}
\gamma_{-i+l+z_1,-i+z_1}\gamma^*_{j+z_2,j+l+z_2}\qudit{e_{-i+l+z_1,-i+z_1}}\rqudit{e_{j+z_2,j+l+z_2}}\Biggr)\Biggr)\nonumber\\
& = \frac{1}{d^2}\sum_{s\in\vert
S\vert}^{}{}\sum_{i=0}^{d-1}{}\sum_{j=0}^{d-1}{}\sum_{l=0}^{d-1}{}\sum_{k=0}^{d-1}{}\Biggl(p_s\alpha^s_{ij}
\qudit{i+l}\rqudit{j+l}\\ &\quad\otimes\Biggl(\sum_{z_1=0}^{d-1}{}
\sum_{z_2=0}^{d-1}{}\omega^{z_1k}
\omega^{z_2k}\gamma_{-i+l+z_1,-i+z_1}\gamma^*_{j+z_2,j+l+z_2}\qudit{e_{-i+l+z_1,-i+z_1}}\rqudit{e_{j+z_2,j+l+z_2}}\Biggr)\Biggr)\nonumber\\
&  =
\frac{1}{d^2}\sum_{l=0}^{d-1}{}\sum_{k=0}^{d-1}{}\Biggl(\sum_{s\in\vert
S\vert}^{}{}\sum_{i=0}^{d-1}{}\sum_{j=0}^{d-1}{}\Biggl(p_s\alpha^s_{ij}
\qudit{i+l}\rqudit{j+l}\\&\quad\otimes\Biggl(\sum_{z_1=0}^{d-1}{}
\sum_{z_2=0}^{d-1}{}\omega^{z_1k}\omega^{z_2k}\gamma_{-i+l+z_1,-i+z_1}
\gamma^*_{j+z_2,j+l+z_2}\qudit{e_{-i+l+z_1,-i+z_1}}\rqudit{e_{j+z_2,j+l+z_2}}\Biggr)\Biggr)\Biggr)\nonumber\\
& =
\frac{1}{d^2}\sum_{l=0}^{d-1}{}\sum_{k=0}^{d-1}{}\Biggl(\sum_{s\in\vert
S\vert}^{}{}\sum_{i=0}^{d-1}{}\sum_{j=0}^{d-1}{}\Biggl(p_s\omega^{ik}\omega^{-jk}\alpha^s_{ij}
\qudit{i+l}\rqudit{j+l}\\&\quad\otimes\Biggl(\sum_{z_1=0}^{d-1}{}
\sum_{z_2=0}^{d-1}{}\omega^{-ik}\omega^{jk}\omega^{z_1k}\omega^{z_2k}
\gamma_{-i+l+z_1,-i+z_1}\gamma^*_{j+z_2,j+l+z_2}\qudit{e_{-i+l+z_1,-i+z_1}}\rqudit{e_{j+z_2,j+l+z_2}}\Biggr)\Biggr)\Biggr)\nonumber\\
&  =
\frac{1}{d^2}\sum_{l=0}^{d-1}{}\sum_{k=0}^{d-1}{}\Biggl(\sum_{s\in\vert
S\vert}^{}{}\sum_{i=0}^{d-1}{}\sum_{j=0}^{d-1}{}\Biggl(p_s\omega^{ik}\omega^{-jk}\alpha^s_{ij}
\qudit{i+l}\rqudit{j+l}\otimes\Biggl(\sum_{z'_2=0}^{d-1}{}
\sum_{z'_2=0}^{d-1}{}\omega^{z'_1k}\omega^{z'_2k}\gamma_{z'_1+l,z'_1}\gamma^*_{z'_2,z'_2+l}\qudit{e_{z'_1+l,z'_1}}\rqudit{e_{z'_2,z'_2+l}}\Biggr)\Biggr)\Biggr)\nonumber\\
&  =
\frac{1}{d^2}\sum_{l=0}^{d-1}{}\sum_{k=0}^{d-1}{}\Biggl(\sum_{s\in\vert
S\vert}^{}{}\sum_{i=0}^{d-1}{}\sum_{j=0}^{d-1}{}\Biggl(p_s\omega^{ik}\omega^{-jk}\alpha^s_{ij}
\qudit{i+l}\rqudit{j+l}\Biggr)\otimes\Biggl(\sum_{z'_2=0}^{d-1}{}
\sum_{z'_2=0}^{d-1}{}\omega^{z'_1k}\omega^{z'_2k}\gamma_{z'_1+l,z'_1}\gamma^*_{z'_2,z'_2+l}\qudit{e_{z'_1+l,z'_1}}\rqudit{e_{z'_2,z'_2+l}}\Biggr)\Biggr)\nonumber\\
&  =
\frac{1}{d^2}\sum_{l=0}^{d-1}{}\sum_{k=0}^{d-1}{}X_lZ_k\Biggl(\sum_{s\in\vert
S\vert}^{}{}\sum_{i=0}^{d-1}{}\sum_{j=0}^{d-1}{}p_s\alpha^s_{ij}
\qudit{i}\rqudit{j}\Biggr){(X_lZ_k)}^\dagger\otimes\gamma_{l,k}\gamma^*_{l,k}\qudit{e_{l,k}}\rqudit{e_{l,k}}\nonumber
\end{split}\\
&&\hskip.5em=
\frac{1}{d^2}\sum_{l=0}^{d-1}{}\sum_{k=0}^{d-1}{}X_lZ_k\rho{(X_lZ_k)}^\dagger\otimes\gamma_{l,k}\gamma^*_{l,k}\qudit{e_{l,k}}\rqudit{e_{l,k}}.\label{nearlydone}
\end{eqnarray}\end{widetext}
Finally, on performing a partial trace over the environment system on
Eq.~\ref{nearlydone}, we obtain, up to a scaling factor,
\begin{eqnarray}\label{last}
\frac{1}{d^2}\sum_{l=0}^{d-1}{}\sum_{k=0}^{d-1}{}X_lZ_k\rho{(X_lZ_k)}^\dagger.
\end{eqnarray}
Equation \ref{last} represents a model for the operator-sum representation of a quantum operation
explicitly in terms of operators on the principal system. We identify the set  $\{{1}/{d}X_lZ_k,
(l,k) \in \Z_d\times\Z_d\}$ as being a possible set of operation elements for such a representation.
\begin{remark}
We note that after performing a partial trace over environment, the state of the system is always expressible
in terms of an operator-sum representation. As such, the  elements $1/d X_lZ_k$ for $
l,k \in \{0,\dots,d-1\}$ do not provided a unique representation for the operation elements for a quantum operation.
\end{remark}

\section{Conclusion}
We explicitly derived a set of elements that act on a $d$-dimensional Hilbert space. We
demonstrated that the derived set forms a basis for the set of bounded operators  acting on
the Hilbert space $\C^d$. As an application, we considered the operator-sum representation of a quantum operation,
and illustrated how the derived set may be identified with a set of operational elements
for such an operation.

\acknowledgements It is a pleasure to acknowledge the help of
Prof.~Peter Wild. The author would also like to thank
Prof.~Matthew G.~Parker,  Prof.~R\"udiger Schack and Dr.~Mazhar
Ali for helpful comments.

\bibliography{Bib}

\end{document}